\documentclass[aps, reprint]{revtex4-2}

\usepackage{graphicx}
\usepackage{amsmath}
\usepackage{natbib}
\usepackage{orcidlink}

\begin{document}

\title{Modelling Soil as a Living System:\\Feedback between Microbial Activity and Spatial Structure}
\author{Riz Fernando Noronha \orcidlink{0009-0007-2923-3835}}
\email{noronha@nbi.ku.dk}
\author{Kim Sneppen \orcidlink{0000-0001-9820-3567}}
\author{Kunihiko Kaneko \orcidlink{0000-0001-6400-8587}}
\affiliation{Niels Bohr Institute, University of Copenhagen, Copenhagen 2200, Denmark}

\begin{abstract}

Soil is a complex, dynamic material, with physical properties that depend on its biological content. We propose a cellular automaton model for self-organizing soil structure, where soil aggregates and serves as food for microbial species. These, in turn, produce nutrients that facilitate self-amplification, establishing a cyclical dynamic of consumption and regeneration. Our model explores the spatial interactions between these components and their role in sustaining a balanced ecosystem. The main results demonstrate that (1) spatial structure supports a stable living state, preventing population collapse or uncontrolled growth; (2) the spatial model allows for the coexistence of parasitic species, which exploit parts of the system without driving it to extinction; and (3) optimal growth conditions for microbes are associated to diverse length scales in the soil structure, suggesting that heterogeneity is key to ecosystem resilience. These findings highlight the importance of spatio-temporal dynamics of life in soil ecology.
    
\end{abstract}

\maketitle

\section{Introduction}

Soil is of fundamental importance to many facets of life: it forms the basis of agriculture, construction, ecosystems, and much more. It provides plants with the environment and necessary nutrients to grow, thus providing the world with food, oxygen, and energy. Over half of the species on earth are estimated to call the soil their home \cite{anthony2023enumerating}, leading to remarkably complex and interesting ecosystems.

Soil is not a physico-chemical material in contrast to sand or rocks, but is shaped through an interplay between organic process and material. Through motility, consumption, and chemical secretion, organisms including a vast number of microorganisms shape the physical structure of soil, whereas such structure affords growth of organisms living in it. Ettema and Wardle \cite{ettema2002spatial} describe how species in soil are neither homogeneously distributed nor random, but instead show spatial heterogeneity and patterning. This spatial heterogeneity plays a role in determining the spatial distribution of plant communities as we observe them above the soil. Soil appears to be a self-organizing substance, with properties that are linked to underlying microbial activity \cite{crawford2012microbial, neal2020soil, feeney2006three}.

While a lot of research has investigated how an ecosystem can affect the soil structure, there remains a paucity of studies exploring co-occurring interactions of how soil structure could affect the ecosystem's survival. It may be possible that species living in soil shape its structure in a way that benefit their continued existence. This paper introduces a simple model to explore the mutual feedback mechanism between physical spatial structure and the microbial ecosystem. A key idea is chemical secretion and cooperation, where a replicating individual needs to feed on nutrients generated by other individuals \cite{baran2015exometabolite, glendsouza2018ecology, embree2015networks}.

\section{Simple Model and Simulation}

Soil consists of multiple different species and chemicals. To simplify, we first develop a model with one top-level organism, one nutrient, and a single soil state. The top-level organism is denoted a microbe, and represents some abstract biological entity that could be a protist or some other slightly larger creature. We consider four states: [S]oil, [N]utrient, [M]icrobe, and [E]mpty. The model can be described with the following dynamics:
\begin{itemize} \itemsep0em
    \item[-] Microbes convert soil sites into nutrient sites
    \item[-] Microbes consume nutrient sites to reproduce
    \item[-] Microbes die at a constant rate $\theta$
    \item[-] Soil sites `grow' into neighbouring empty and nutrient sites at rate $\sigma$
\end{itemize}

The soil `growth' can be justified by considering soil as a living material that includes soil aggregation, as noted by Tisdall and Oades \cite{tisdall1982organic}. In this language, one may think of empty and nutrient states as sites containing `dust' of non-aggregated inert matter.

\begin{figure}
    \centering
    \includegraphics[width=\linewidth]{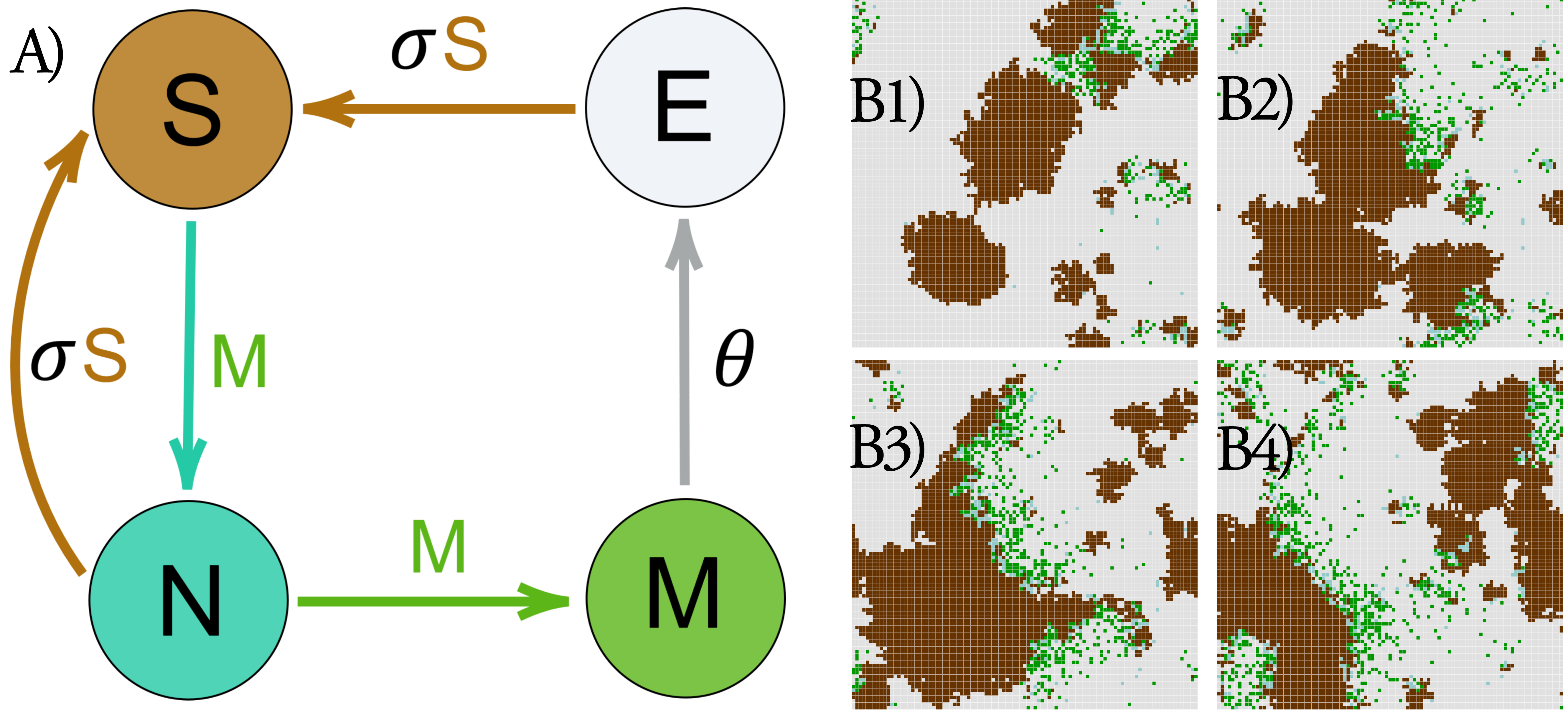}
    \caption{A): Schematic diagram of the model. S,E,N,M represent Soil, Empty, Nutrient, and Microbe states respectfully. Arrows indicate transitions, and the text above them the rates. B): Snapshots of the 2D lattice at different time steps, showing dynamic spatial behaviour.}
    \label{fig:SENM_schematic}
\end{figure}

By disregarding spatial heterogeneity, the system can be described with the following mean-field equations:

\begin{align}
    \frac{\mathrm{d}S}{\mathrm{d}t} &= \sigma S \left(E+N\right) - MS \label{eq:dSdt_nutrient} \\
    \frac{\mathrm{d}E}{\mathrm{d}t} &= \theta M - \sigma SE \label{eq:dEdt_nutrient} \\
    \frac{\mathrm{d}N}{\mathrm{d}t} &= MS - MN -\sigma SN \label{eq:dNdt_nutrient} \\
    \frac{\mathrm{d}M}{\mathrm{d}t} &= MN - \theta M \label{eq:dMdt_nutrient}
\end{align}

If we add Equations \ref{eq:dSdt_nutrient}, \ref{eq:dEdt_nutrient}, \ref{eq:dNdt_nutrient}, and \ref{eq:dMdt_nutrient}, we get
\[ \frac{\mathrm{d}S}{\mathrm{d}t} + \frac{\mathrm{d}E}{\mathrm{d}t} + \frac{\mathrm{d}N}{\mathrm{d}t} + \frac{\mathrm{d}M}{\mathrm{d}t} = 0 \]
\begin{equation}
    S + E + N + M = \mathcal{C} \label{eq:conservedSENM}
\end{equation}
where $\mathcal{C}$ is a constant, determined by the initial conditions. This is a 3 variable system, as a fourth variable can be easily eliminated through \autoref{eq:conservedSENM}. In addition, this introduces a carrying-capacity set by $\mathcal{C}$ into the system: if the system is composed of mostly soil, then further growth of soil in the system ($\sigma S E$) goes as $\sim S (\mathcal{C}-S)$, which corresponds to a standard Verhulst or logistic equation $\mathrm{d}S/\mathrm{d}t\approx S(1-S)$ \cite{verhulst1838notice}. The same carrying capacity acts on the microbe, as $MN\sim M(\mathcal{C}-M)$.

The model can thus be thought of as a modified predator-prey model \cite{lotka1920analytical}, with a time delay (as the microbes consume soil in a two-step process: first converting it to nutrients, and then eating them to reproduce) and a carrying capacity. In particular, without the nutrient step (if the microbe directly consumes soil to survive), we see no sustained oscillations. Furthermore, the phase diagram would have no bifurcations, and the fraction of microbes smoothly varies as we change the parameters $\sigma$ and $\theta$.

\begin{figure}
    \centering
    \includegraphics[width=\linewidth]{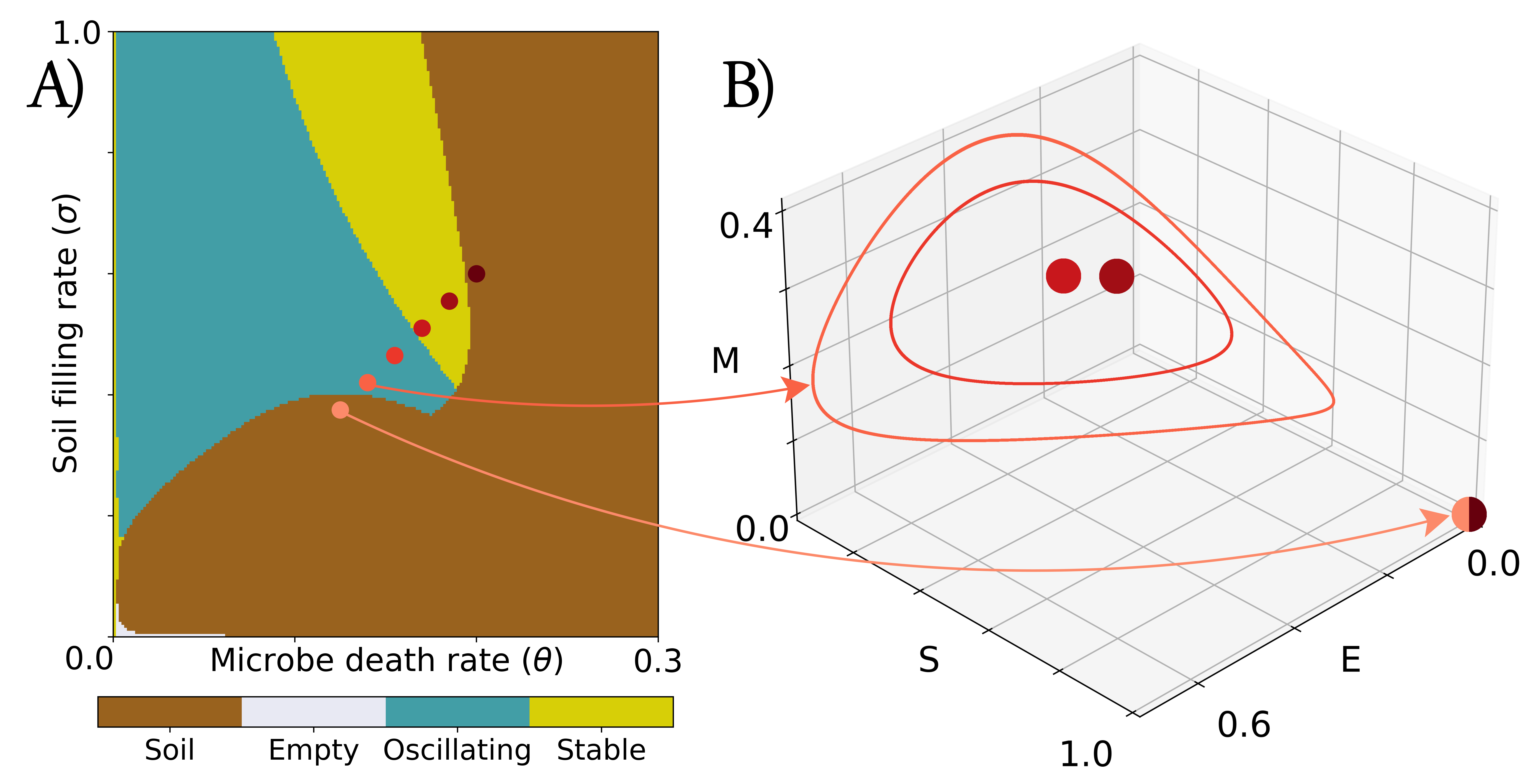}
    \caption{A): Phase diagram of the attractors of the mean-field model over soil filling rate $\sigma$ and microbe death rate $\theta$. B): Attractors of the corresponding red points in A), in state space $S$, $E$, $M$ represent Soil, Empty and Microbe populations. Fixed points are represented by circles, while limit cycles (oscillations) can be seen as closed loops.}
    \label{fig:meanfield_statespace}
\end{figure}

Using the above equations, we can construct a mean-field phase diagram, visualized in \autoref{fig:meanfield_statespace}. The phase diagram contains three bifurcations: a Hopf bifurcation boundary, where the stable limit cycle changes to a stable fixed point, a saddle-node bifurcation on the right, beyond which the system dies out, and a saddle-node on invariant curve (SNIC) bifurcation at the bottom, which again makes survival impossible. Below the SNIC bifurcation, we can observe that the oscillations have too large an amplitude, which leads to the system falling into the pure soil absorbing state \footnote{Due to the bifurcations, we can section the phase diagram into four distinct regimes. Note that two fixed points exist everywhere, namely the pure soil absorbing state (as without microbes, soil cannot be destroyed) and the pure empty absorbing state (as without soil, empty states will only proliferate). The first is the area to the right, where there are no fixed points beyond the two trivial ones. Then, we come across the saddle-node bifurcation, which introduces two fixed points: an unstable saddle point, and a stable fixed point (which allows for species survival). Crossing the Hopf bifurcation causes the stable fixed point to swap stability and become unstable, but creates a stable limit cycle. Finally, crossing the SNIC boundary to the bottom causes the stable limit cycle to collide with a nullcline and fall into the trivial pure-soil state, and thus we have only an unstable fixed point in the system, causing it to die out.}.

Apart from analyzing the mean-field equations, we perform numerical simulations on a 2-dimensional lattice in order to investigate the impacts of spatial structure on the model. In our lattice, each site holds a certain state: either soil, empty, nutrient or microbe. A site cannot hold more than one state at a time. The lattice is updated through a stochastic sequential update algorithm, in which we first choose a random site, and perform some dynamics based on the site's value. If it is empty or a nutrient, we do nothing; if it is soil, we choose a random neighbour, and if the neighbour is either empty or nutrient, we change the neighbour to soil with rate $\sigma$; and if it is a microbe, we first check for death (with rate $\theta$), and if it does not die we move it to a random adjacent site. If the adjacent site was a soil site, it leaves behind a nutrient, and if the adjacent site was a nutrient, leaves a copy of itself behind it.

\begin{figure}
    \centering
    \includegraphics[width=\linewidth]{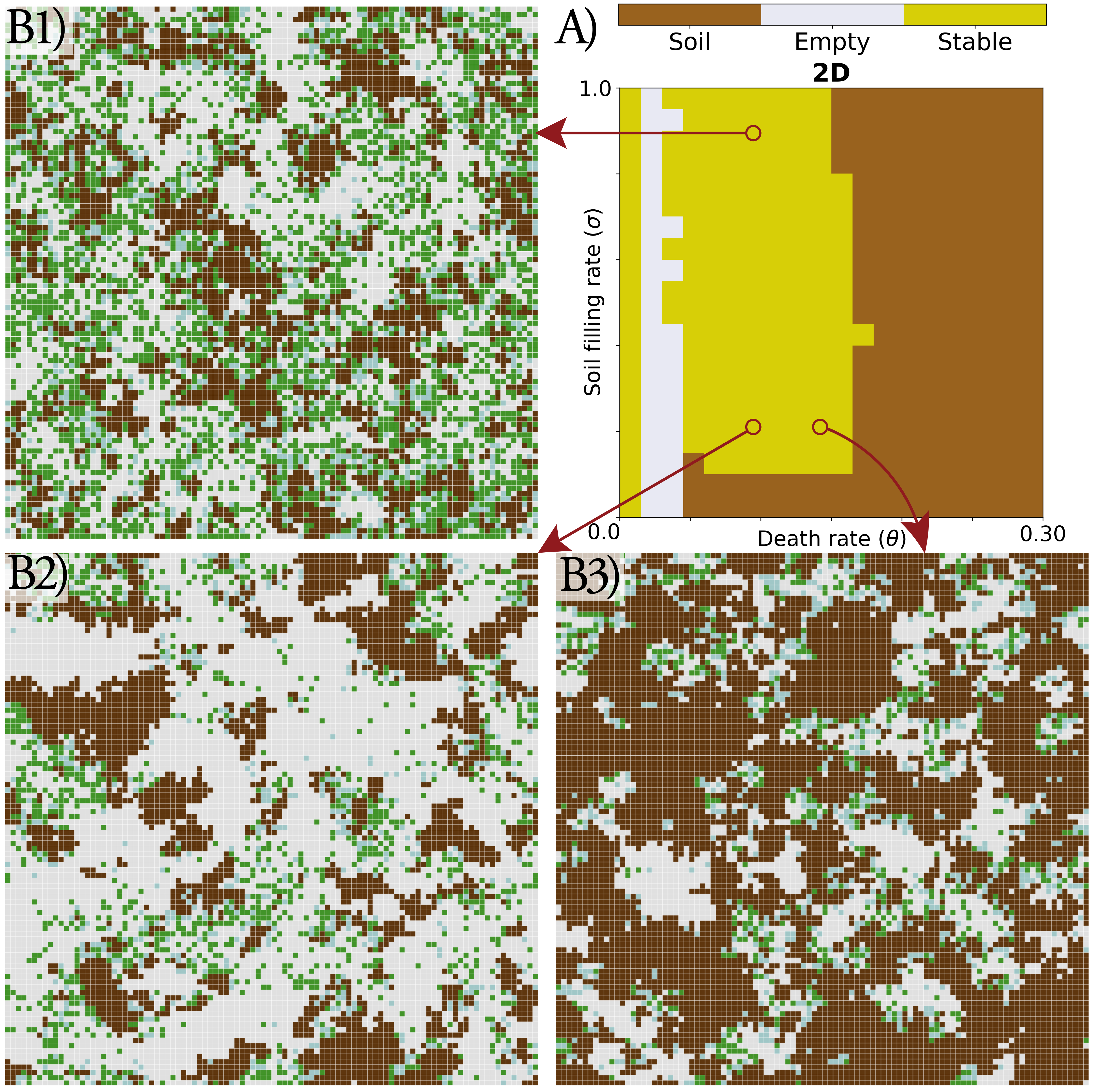}
    \caption{A): Phase diagram over soil filling rate $\sigma$ and microbe death rate $\theta$ for a 2D lattice. B) Snapshots of the long-term behaviour on the lattice for different parameter values.}
    \label{fig:2D_4params}
\end{figure}

Two parameters, namely the soil-filling rate $\sigma$ and the microbe death rate $\theta$, control the equilibrium state of the system. Increasing $\sigma$ leads to an increase in microbe states, similar to standard predator-prey models where increasing the prey's growth rate leads to an increase in predator population. Increasing $\theta$ leads to an increased amount of soil in the system. \autoref{fig:2D_4params} shows the phase diagram, as well as the snapshot pattern. The snapshots indicate that the microbes prefer to stay near the surfaces of `aggregations' of soil, which appears to coincide with experimental evidence \cite{wilpiszeski2019soil}.

The two dimensional system is observed to have local oscillations (\autoref{fig:SENM_schematic} B) shows evidence of traveling waves with two clear wavefronts: a soil wavefront expanding into empty regions, and a microbe wavefront expanding into the soil). However, as the system size increases, the oscillations become decoherent in space, with regions of space oscillating out of phase, and as such the population counts of the system as a whole are not oscillatory. Through the lack of coherent oscillations, the system can survive in a parameter regime where it dies in the mean-field (below the SNIC bifurcation). Here the microbes support each other through produced nutrients so that spatial structure is beneficial for their survival.

\section{Soil Particle Size Distribution}

\begin{figure}
    \centering
    \includegraphics[width=\linewidth]{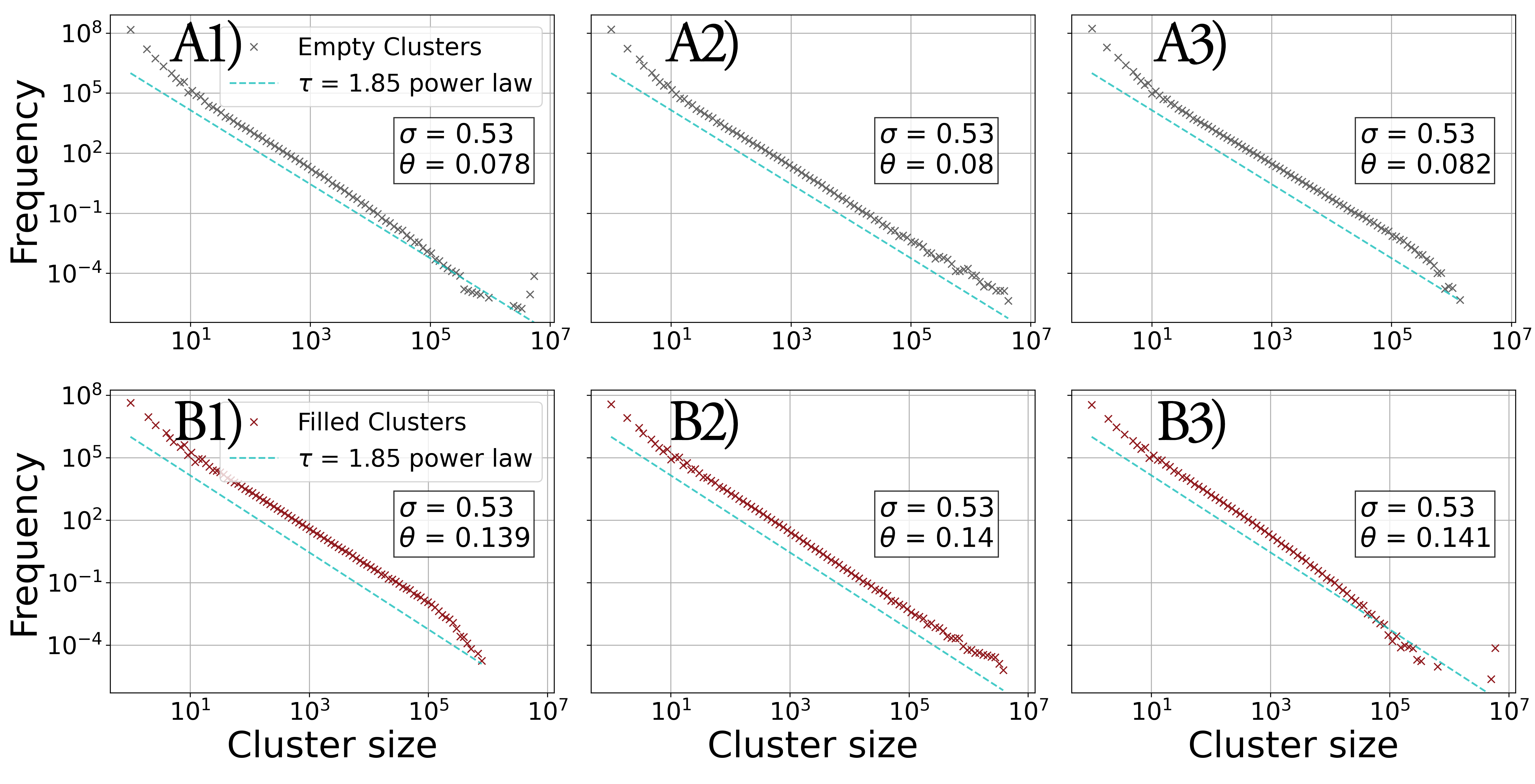}
    \caption{Cluster size distribution calculated on a 2D lattice with soil filling rate $\sigma$=0.53, $L$=4096, for increasing microbe death rate $\theta$. A): Empty site cluster size distributions: supercritical (A1), critical (A2), and subcritical (A3). B): Filled cluster size distributions, subcritical (B1), critical (B2) and supercritical (B3).}
    \label{fig:powerlaw}
\end{figure}

We next investigate soil and empty cluster sizes. In our simulations, this is done by defining the nearest-neighbour connected clusters of soil sites as a `soil particle', and likewise for empty clusters. We observe that by fixing the soil filling rate $\sigma$ and gradually increasing the microbe death rate $\theta$ allows us to observe two critical points: one where the empty cluster size distribution follows a power law, and one where the soil cluster size distribution follows a power law (\autoref{fig:powerlaw}). Soil particles are observed to have a power law distribution in nature \cite{anderson1997applications} and the fact that vacancies in soil could have a fractal structure appears to also align with some observations \cite{ozhovan1993fractal}.
both of the observed cluster size distributions appear to follow a power law, with an exponent of $\approx 1.85$. The formation of soil clusters, by growing into neighbours, can be mapped to the directed percolation universality class and thus we observe something analogous to supercritical directed percolation in 2+1 dimensions \footnote{We can imagine a simplified model with two states, soil and empty. Soil sites grow into their empty neighbours (such as the Eden model \cite{eden1961two}), and die randomly. While soil deaths are not random in our model (they depend on microbes, whose locations are correlated to the surface of the soil) the power-law behaviour appears to be similar. These dynamics describe directed percolation \cite{grassberger2006tricritical}, and coarse-graining supercritical directed percolation beyond the correlation length leads to site percolation}. More extensive simulations on a simplified model revealed that the exponent observed was the same as that of site percolation, namely, an exponent of $\approx 2.05$.

\begin{figure}
    \centering
    \includegraphics[width=\linewidth]{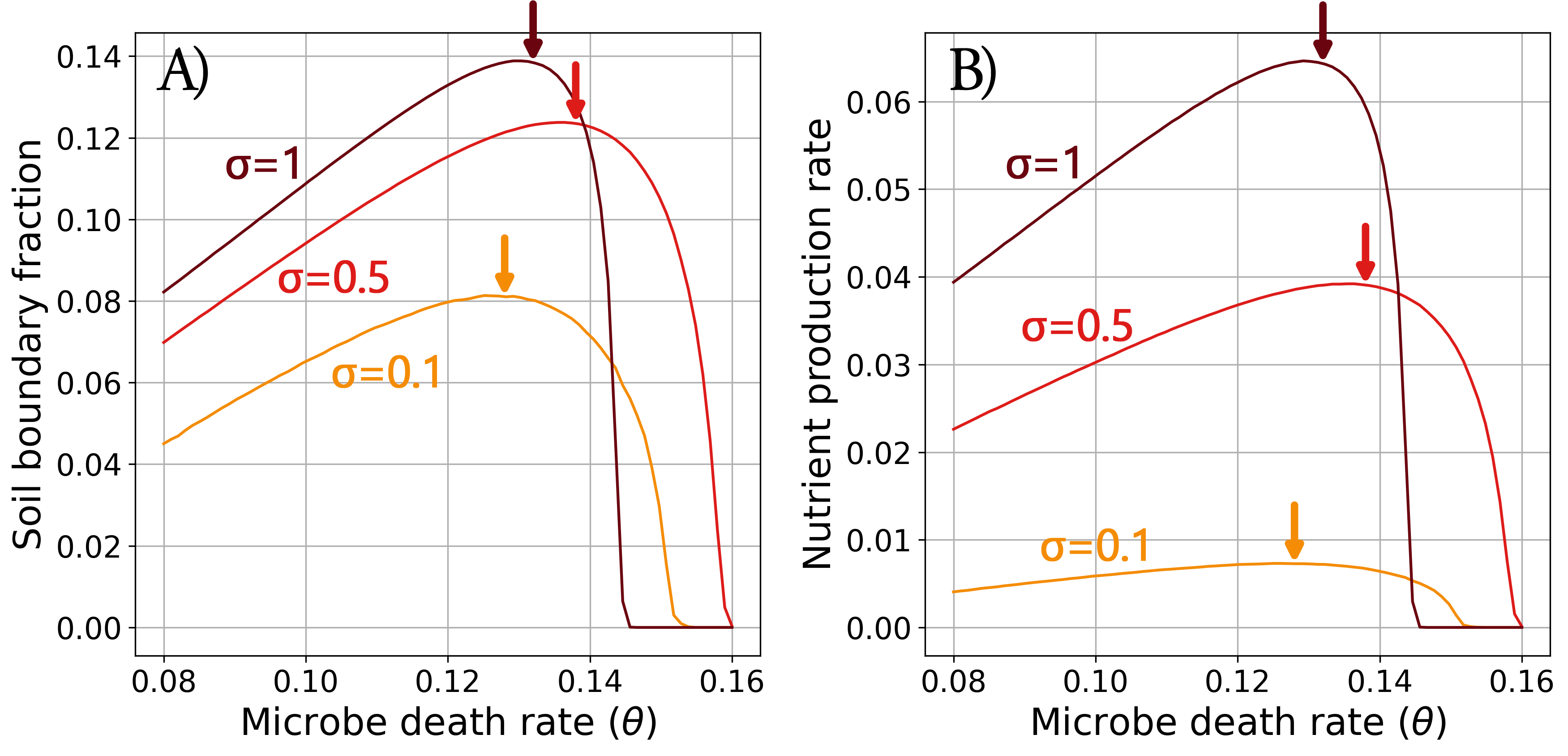}
    \caption{Plots of soil boundary fraction,i.e, the number of soil boundaries, normalized by the maximum possible number (A) and nutrient production rate (B) vs microbe death rate $\theta$ for different values of soil-filling rate $\sigma$ show a correlation between the soil power-law point (marked with an arrow), the nutrient production maximum, and the soil boundary maximum.}
    \label{fig:nutrient_maximization_power_law}
\end{figure}

The location of the soil power-law was observed to correspond with the parameter values that maximize the nutrient-production in the system (\autoref{fig:nutrient_maximization_power_law}). In particular, the soil power-law is correlated with a maximization of soil boundaries (in particular, the number of places where a soil site borders a non-soil site). The soil boundary maxima can be seen to intuitively correspond to the nutrient production maximization from \autoref{eq:dNdt_nutrient}: more boundaries make it easier for microbes to create nutrients. Too dense blocks of soil (the supercritical state) have fewer boundaries, as a non-porous structure is increased, while too little soil (the subcritical state) lacks the amount of soil sites needed to create a high number of soil boundaries.

\section{Two Species Modelling}

As real soil contains a diverse number of species, we extend the model by adding in a second species of microbe. We study two cases: symbiotic and parasitic; in the former two species support each other by creating the nutrients required by the other, whereas in the latter the parasite exploits the nutrient that is created by the host species.

\begin{figure}
    \centering
    \includegraphics[width=\linewidth]{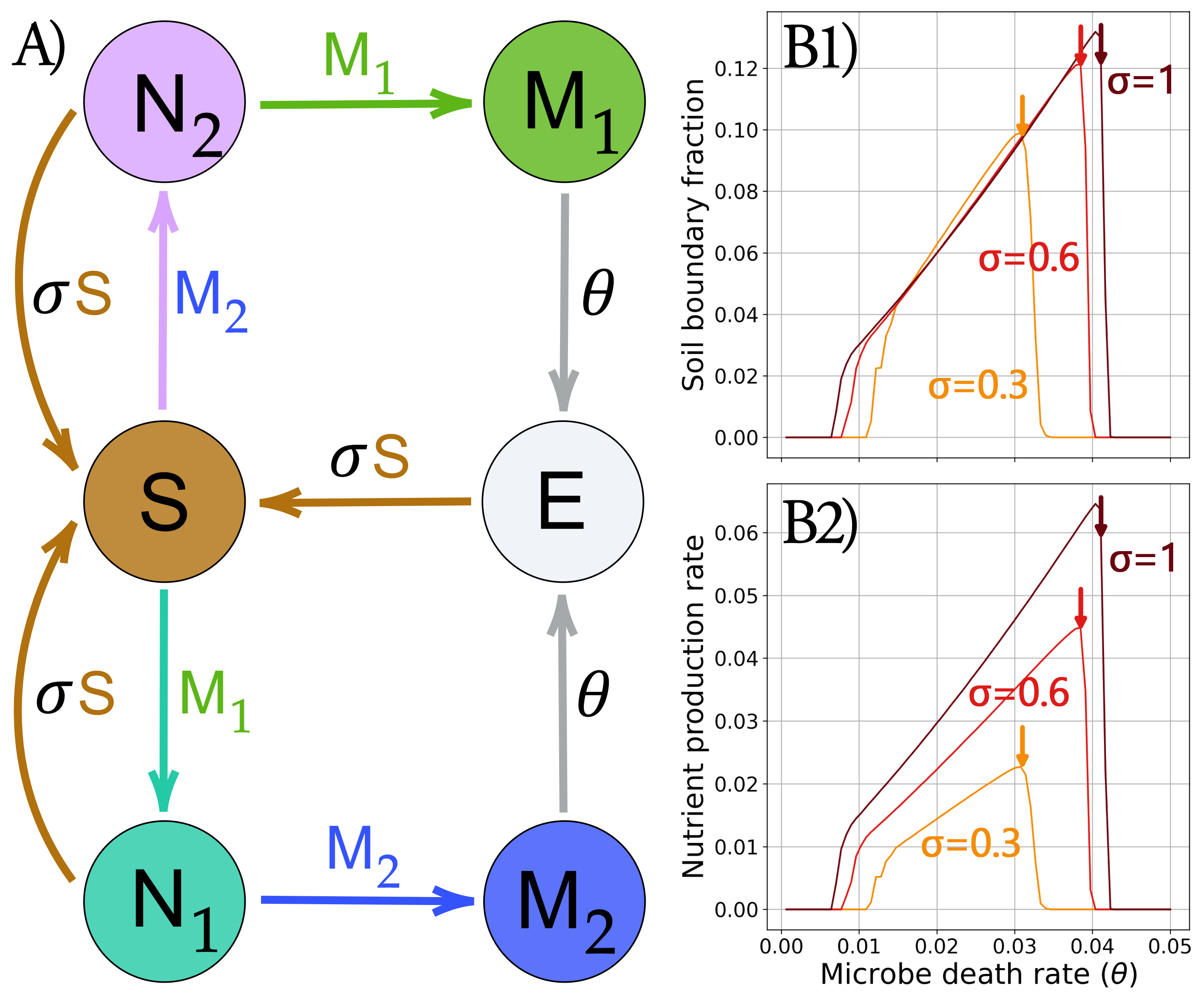}
    \caption{A): Schematic diagram of the symbiotic model. Aside from Soil (S) and Empty (E) we now have two microbes (M1 and M2) with their own nutrients (N1 and N2). Microbe 1 needs the nutrient created by Microbe 2 to survive, and vice versa. Arrows indicate transitions, and the text above them the rates. B): Plots of soil boundaries (B1) and nutrient production (B2) vs microbe death rate $\theta$ for different values of soil-filling rate $\sigma$ show a correlation between the soil power-law regime (marked with an arrow), the nutrient production maximum, and the soil boundary maximum.}
    \label{fig:twospec_schematic}
\end{figure}

\textbf{Symbiotic Case:} We set up the system so that the species need to mutually support each other: microbe 1 needs to eat the nutrient created by microbe 2 to reproduce, and vice versa. The fact that microbes now create nutrients that are useless to themselves means survival is harder than the single-species model. Survival requires both microbes to be present and near each other, to cross-feed, and thus the spatial structure hurts the system, making it less stable than the mean field. As such, we observe that the survival region shrinks as compared to the single species case. However, we also observe that even in the restricted coexistence region, the nutrient production maximization (which is observed to be sharper) is correlated with the soil power-law (\autoref{fig:twospec_schematic}).

In addition, we simulated models with multiple species in a cyclical autocatalytic network, where species $i$ consumes the nutrient of $i+1$, and so on. While the survival region shrank, we still observed that coexistence was possible, in a network of more than four species, and spiral-wave-like patterns were observed, similar to previous studies that used more classical catalysis without the intermediate nutrient step \cite{takeuchi2012evolutionary}. Furthermore, in simulations of up to 6 species, we observed that the power-law was still obtained within the shrunken survival region.

\textbf{Parasitic Case:} To study the behavior in the presence of parasites, we modify the model so that the second species cannot create its nutrient (when it moves into a soil site, it simply switches places\footnote{We also made a model where the parasite destroyed the soil, and left behind an empty site. Both had similar results, however, in the latter model, the parasite suppressed the hosts further: not only did it steal their nutrients, but also destroyed soil that would have potentially served to create nutrients. As such, the resulting system was observed to be more soil-dense than the former.}), but still consumes the nutrients of the first species to reproduce. The parasitic organism does not need the cost to produce nutrients, so we give it an advantage in the form of a lower death rate $\theta$. {The following set of equations can now describe the mean-field dynamics:

\begin{align}
\frac{\mathrm{d}S}{\mathrm{d}t} &= \sigma S (E + N) - S M_1 \\
\frac{\mathrm{d}E}{\mathrm{d}t} &= N ( M_1 + M_2) + \theta_1M_1+\theta_2M_2 - \sigma S E \\
\frac{\mathrm{d}N}{\mathrm{d}t} &= S M_1  - N(M_1 + M_2) - \sigma S N \\
\frac{\mathrm{d}M_1}{\mathrm{d}t} &= M_1 N - \theta_1 M_1 \\
\frac{\mathrm{d}M_2}{\mathrm{d}t} &= M_2 N - \theta_2 M_2
\end{align}

Most of the dynamics are similar to the original model (Equations \ref{eq:dSdt_nutrient}, \ref{eq:dEdt_nutrient}, \ref{eq:dNdt_nutrient}, and \ref{eq:dMdt_nutrient}), with the following key differences:
\begin{itemize} \itemsep0em
    \item[-] We have two species of microbes ($M_1$ and $M_2$) with different death rates $\theta_1$ and $\theta_2$
    \item[-] Only $M_1$ contributes to nutrient creation
\end{itemize}

\begin{figure}
    \centering
    \includegraphics[width=\linewidth]{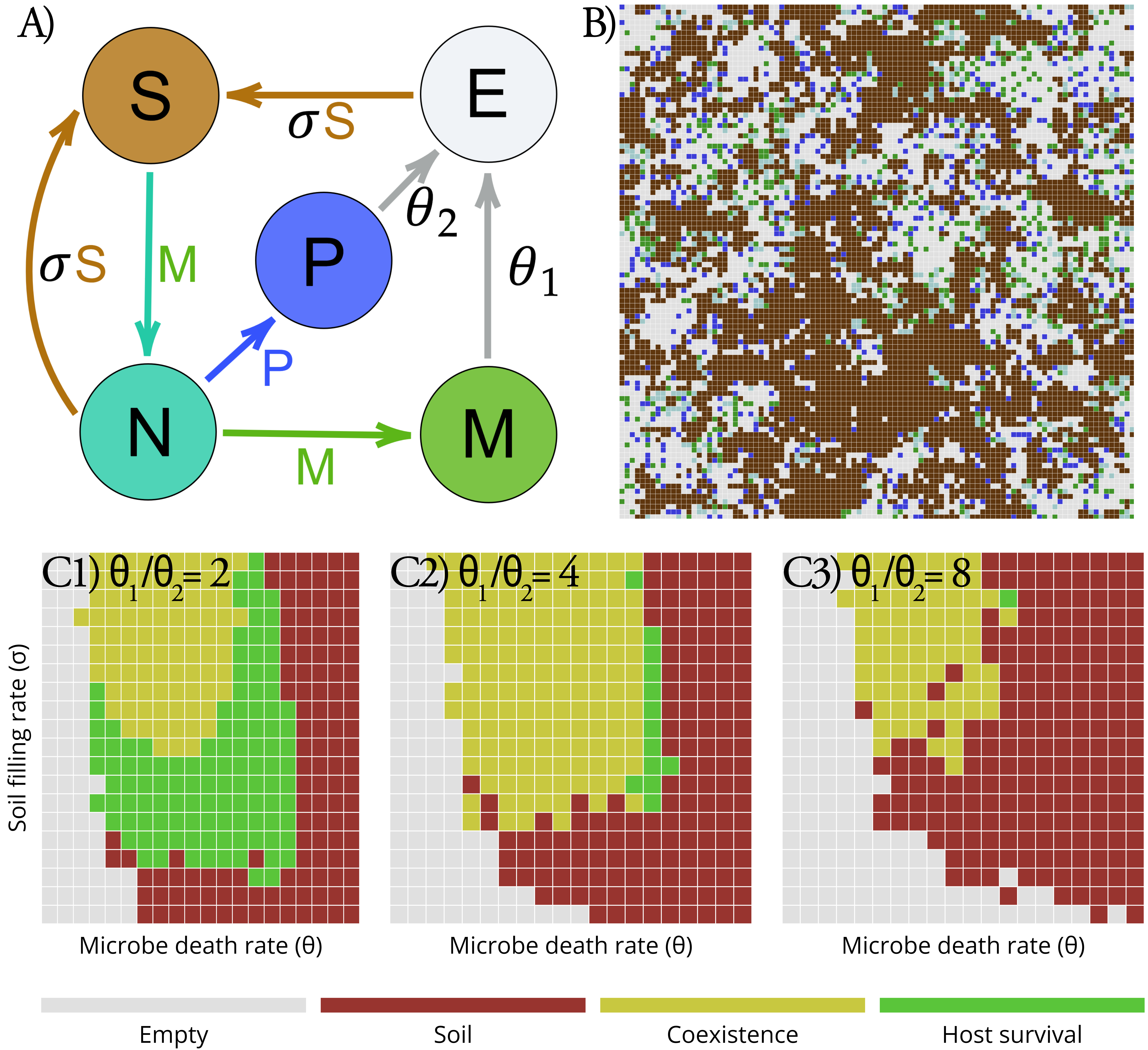}
    \caption{A): A schematic diagram of the rates and transitions in the parasite model. B): A snapshot of the 2D lattice, showing long-term coexistence between hosts (green) and parasites (blue). C): A rasterscan over soil filling rate $\sigma$ and microbe death rate $\theta$ showing parasite coexistence as a function of the parasite's relative longevity, $\theta_1/\theta_2$. C1) $\theta_1/\theta_2$=2; C2)  $\theta_1/\theta_2$=4; C3) $\theta_1/\theta_2$=8. While the total survival region appears to shrink, the ratio of the coexistence area to the total survival area in phase space appears to increase.}
    \label{fig:parasite_coexistence}
\end{figure}

In the mean-field equations, the parasite and host can never coexist: due to the parasite's longer lifespan, by competitive exclusion it outcompetes the host, and then dies as it cannot sustain itself. However, parasite-host coexistence was observed on a 2D lattice, similar to previous studies \cite{stump2018local}. The host species create nutrients around themselves, and thus despite having a higher death rate, can find the nutrients and reproduce more easily than the parasites, which have to randomly walk until they find nutrients they can use to reproduce \footnote{The role of spatial structure in the avoidance of extinction by parasites in a catalytic reaction network was also reported by \cite{boerlijst1991spiral, altmeyer2001error}}.

We observe that at low relative lifespans $\theta_1/\theta_2$ (corresponding to the parasite living only slightly longer than the host), there are large regions of parameter space where the parasite cannot survive and goes extinct (\autoref{fig:parasite_coexistence}). By making the parasite more effective by increasing its effective lifespan $\theta_1/\theta_2$, we observe that while the total range of parameters in which the host survives goes down, the coexistence region occupies a greater fraction of the entire survival region.

However, a higher parasite advantage leads to more soil in the system (as the presence of the parasite suppresses the host), and for extremely large values of $\theta_1/\theta_2$, the particle size distribution of soil clusters is supercritical regardless of parameters. Even in cases where we can observe power-law distributed particle sizes, the power law no longer corresponds to the nutrient production maximum.

\section{3D Model}

We can generalize the same stochastic sequential algorithm to perform simulation on a 3-dimensional lattice. In 3 dimensions, we observe coherent oscillations across the whole system, similar to the mean-field case \footnote{It is possible that the coherent oscillations observed are an effect of a small system size. However, the amplitude of the oscillations did not appear to significantly decay by increasing the system size from $L=40$ to $L=75$.}. Similar to the 2D case, the 3D system is able to survive at parameters where the mean-field system dies, in particular below the SNIC bifurcation. This can be seen to be due to oscillations: in the mean-field, oscillation amplitudes get too large and drive the system to the pure-soil absorbing state. However, in three dimensions, the spatial structure effectively damps the oscillations and thus allows for survival (\autoref{fig:3d_mf_timeseries}).

\begin{figure}
    \centering
    \includegraphics[width=\linewidth]{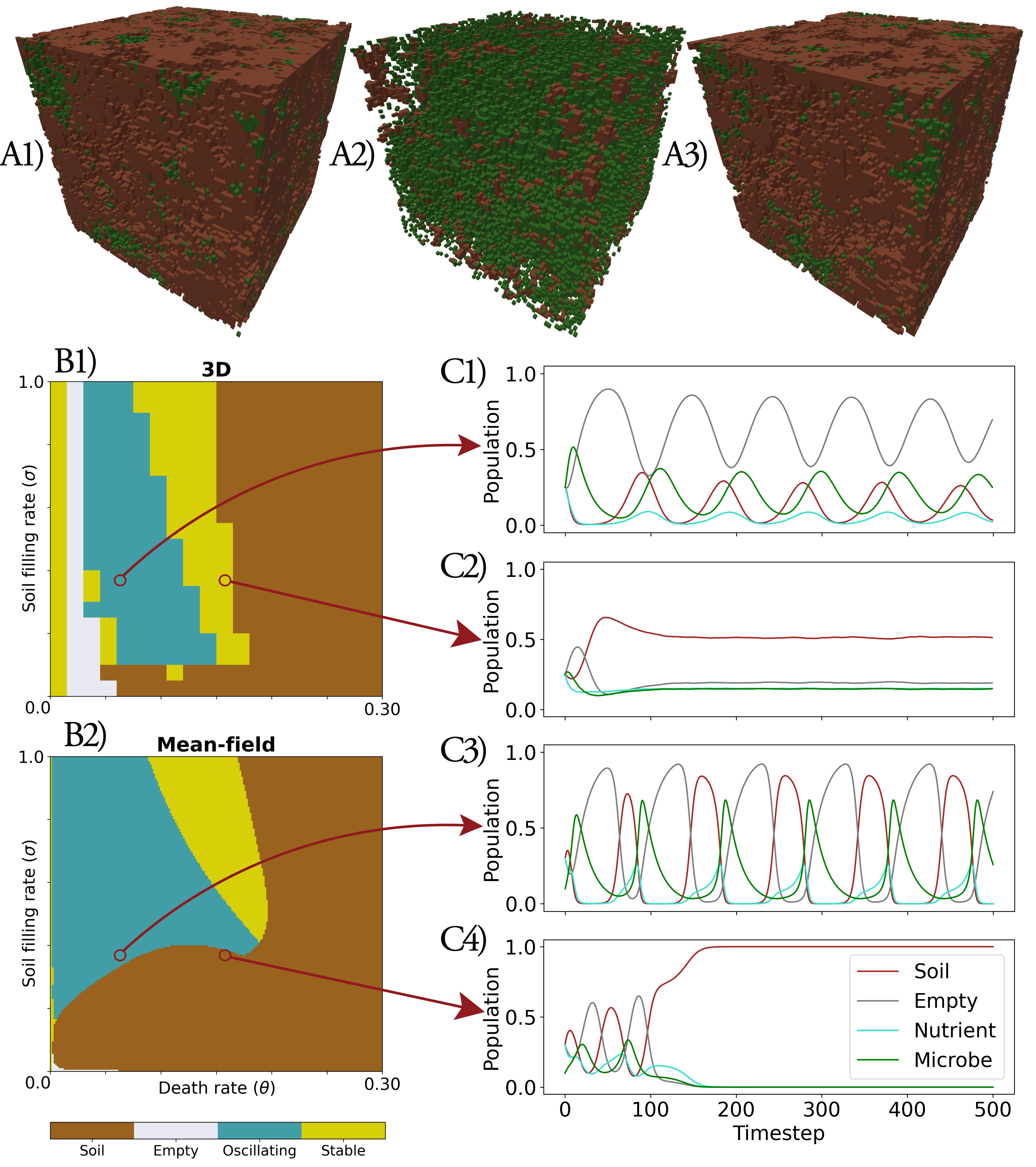}
    \caption{A): Three sequential snapshots of the 3D lattice ($L=75$) with only soil and microbe states visualized, showing oscillations between a soil-dense state and a microbe/empty dense state. B): Phase diagrams over soil-filling rate $\sigma$ and microbe death rate $\theta$ for a 3D lattice ($L=75$) (B1) and mean-field (B2, reproduced from \autoref{fig:meanfield_statespace} for comparison). C): Timeseries for different parameter values.}
    \label{fig:3d_mf_timeseries}
\end{figure}

The coherent oscillations in the 3-dimensional system make it hard to define an exact `critical point' where we observe power-law behaviour. Typically, we observe the amount of soil in the system oscillating between soil-rich and soil-poor states. In soil-rich states, the system appears supercritical, while in soil-poor states it appears subcritical. While we can attempt to find a `critical point' by scanning over the instability and averaging out the oscillations to look for a power-law, the process is not ideal. Furthermore, 3-dimensions makes it easier for a soil site to be connected to an infinite cluster, which leads to the `power-law' occurring very quickly at low microbe death rates \footnote{The order of the empty and soil power laws are swapped in 3D. Increasing the death rate, we first observe the filled `power law' (although oscillations make it hard to call it a power law) and next come across the empty power law. This `swapping' of critical points from 2D to 3D can be understood by percolation theory. In standard site percolation, a site is randomly chosen to be `active' or `inactive' with a probability $p$. Thus, if active sites percolate at probability $p_c$, inactive sites percolate at $(1-p_c)$. In two dimensions, $p_c\approx0.59$ \cite{newman2000siteperc2D}, while in 3D, $p_c\approx0.31$ \cite{lorenz1998siteperc3D}. This `movement' of the critical threshold across $0.5$ leads to the swap in critical points.}.

Once again, similar to the 2D case, the nutrient production maximization is correlated with the soil-boundary maximization. However, due to the high connectedness of 3D, this point is not close to the soil-power law (which is found in the oscillatory regime), but somewhat in the vicinity of the empty site power law (which is found beyond the bifurcation boundary, where the system does not appear to be globally oscillating).

\section{Discussion}

To explore the interplay between soil structure formation and organismal process, we presented a simple model consisting of soil, nutrient, microbe, and empty sites. The organisms create nutrients, which are used to support the growth of the same or other microbes. Within a certain range of parameters, the microbes survive by forming a spatial soil structure, where spatial localization of microbes enhances their survival, in contrast to the corresponding mean-field model. This is because of spatially induced positive feedback, where microbes support each other's amplification locally while leaving separate regions in ``peace" to restoration by ``soil" regrowth. Thus our spatial model exhibits self-organized lying in fallow, supporting microbial coexistence. In a two-dimensional lattice model, the soil shows power-law spatial structure at around the parameter value where the maximal nutrient production is achieved. The dynamics are oscillatory in 3 dimensions, and oscillatory for a wider range in the mean field, which reduces the region of microbe survival.

The present model involves the two-step processes to convert soil to nutrients and then to its utilization for the microbe growth. This is in part analogous to a predator-prey system where the prey is sessile, and where the predator needs to eat the prey two times to replicate. The two-step reproduction creates an opportunity for the evolution of a parasite that can skip a step and focus on consumption. Noticeably, the parasite  always devastates the system in an unstructured environment but robustly coexists with its ``host" otherwise.

Our model repeatedly highlights the interplay between spatial structure and surviving diversity. While a power-law distribution for soil particle size is a highlight of the generated spatial structure here, it is observed only at particular parameter values, corresponding to the percolation point. However, those same parameters appear to be close to parameters that maximize nutrient production in the system. This indicates the importance of a wide range of soil aggregate sizes to the maintenance of healthy thriving soil \cite{wilpiszeski2019soil}. 

For a symbiotic case where two microbial species mutually support each other, we once again observe that the largest nutrient production occurs at conditions where soil aggregates exhibit a power law scaling. Coexistence of a higher number of symbiotic species is possible, in a narrower parameter regime, overlapping with a power law distribution of soil clusters. This  suggests an interplay between organismal symbiosis of diverse species and the observed scale-free features of soil aggregates\cite{zhang2016symbiosis}.

The power-law observed in our model has exponents identified from site percolation, while the power law in real soil can vary. Furthermore, the percolation analogy is only true on large scales while short-range correlations will make perceived scaling different \cite{xu2021SRMlengthscales}. Real soil has also been observed to not follow an exact power law \cite{fu2022fractal} which has led to multifractal descriptions \cite{perfect1993multifractal, bird2006multifractal, gao2021multifractal, posadas2001multifractal, xia2020multifractal, wu2021multifractal}. This limits the possibility for a quantitative comparison with the result of the present model.

Noticeably, the predicted nutrient production maximization and soil scaling do not coincide in our 3D model. Further, the coherent oscillations in the 3D model would be hard to  imagine in extended soil, which makes us believe our isotropic 3D model fails as a model for soil. Real soil is far from isotropic: the vertical direction is significantly different, due to the direction imposed by vertical penetration of water, oxygen, and other compounds. A future exploration of a 2+1 dimensional model might be necessary.

Parasites are observed to kill the spatially unstructured system represented by the mean-field model, yet coexistence is greatly enhanced when spatial separation is allowed. The presence of a localized structure allows hosts to survive from parasitic attacks \cite{stump2018local, pande2016privatization}, similar to the robustness of hypercycles to parasites explored by Boerlijst and Hogeweg \cite{boerlijst1991spiral}. Power laws imply long range (global) correlations, and thus we hypothesize that the soil power law is not optimal for coexistence. This could explain why nutrient maximization occurs at different parameters.

While spatial microbial models have been used in past studies to demonstrate the effects of spatial structure \cite{pande2016privatization, hoek2017emergence, germerodt2016pervasive, van2021population}, soil (with production and consumption of nutrients) has the added caveat of the ability of the ecosystem to mould the space around it. We introduced a simple microbial exchange model for soil structure to illustrate how life and production of nutrients can affect its environment, and benefit itself. The model allows for extended coexistence despite poor environmental growth rates (small $\sigma$) or emergence of cheaters. We hope that our simple toy model inspires future research into the interplay between the many scales of life and how organisms mutually self-organize and shape the physical environment that supports their existence.

\section{Acknowledgements}

We thank Namiko Mitarai and Thomas Bohr for their helpful discussions. RFN and KK are supported by the Novo Nordisk Foundation Grant No. NNF21OC0065542.

\bibliography{references}

\end{document}